\documentclass[twocolumn,prl,psfig,superscriptaddress]{revtex4}

\usepackage{graphicx}

\begin{document}
\title{Electron spin resonance in the $S$=1/2
quasi-one-dimensional antiferromagnet with Dzyaloshinskii-Moriya
interaction BaCu$_2$Ge$_2$O$_7$ }

\author{S. Bertaina $\S$}
\affiliation{Laboratoire Mat{\'e}riaux et Micro{\'e}lectronique de
Provence,  associ{\'e} au CNRS, Case-142,  Universit{\'e}
d'Aix-Marseille III, 13397 Marseille Cedex 20, France}

\author{V. A. Pashchenko}
\affiliation{ Physikalisches Institut, Johann Wolfgang Goethe
Universit{\"a}t, FOR412, Postfach 111932, 60054 Frankfurt am Main,
Germany}

\author{A. Stepanov}
\affiliation{Laboratoire Mat{\'e}riaux et Micro{\'e}lectronique de
Provence, associ{\'e} au CNRS, Case-142,  Universit{\'e}
d'Aix-Marseille III, 13397 Marseille Cedex 20, France}

\author{T. Masuda $^\dag$}
\affiliation{Department of Advanced Materials Science, The
University of Tokyo, 5-1-5 Kashiwa-no-ha, Kashiwa 277-8581, Japan}
\author{K. Uchinokura $^\ddag$,}
\affiliation{Department of Advanced Materials Science, The
University of Tokyo, 5-1-5 Kashiwa-no-ha, Kashiwa 277-8581, Japan}
\date{\today}
%\maketitle

\begin{abstract}

We have investigated  the electron spin resonance (ESR) on single
crystals of BaCu$_2$Ge$_2$O$_7$ at temperatures between 300 and 2
K and in a large frequency band, 9.6 -134 GHz, in order to test
the predictions of a recent theory, proposed by Oshikawa and
Affleck (OA), which describes the ESR in a  $S$=1/2 Heisenberg
chain with the Dzyaloshinskii-Moriya interaction. We find, in
particular, that the ESR linewidth, $\Delta H$, displays a  rich
temperature behavior. As the temperature decreases from
$T_{max}/2\approx $ 170 K  to 50 K, $\Delta H$ shows a rapid and
linear decrease, $\Delta H \sim T$. At low temperatures, below 50
K, $\Delta H$ acquires a strong dependence on the magnetic field
orientation and for $H \| c$ it shows a $(h/T)^2$ behavior which
is  due to an induced staggered field $h$, according to OA's
prediction.

\end{abstract}
\maketitle

While Bethe's solution \cite{bethe} for a $S$=1/2 Heisenberg
antiferromagnetic chain (1/2HAFC) has been known for more than 70
years, chain-like systems continue to  attract considerable
interest. One of the reasons is that the spin-singlet ground state
of 1/2HAFC,
% characterizing by the gapless quantum-critical regime,
in which quantum fluctuations manifest themselves in an extremely
strong way as $T\rightarrow 0$, is unstable with respect to
perturbations breaking the chain uniformity.
%The best known
%example of such  behavior is  a magnetoelastic instability leading
%to  the dimerization of the  ground state in spin-Peierls
%materials.
This significantly affects the spin excitation spectrum
in the low-energy sector, giving rise to a  gapped state
\cite{hase}.

It appears that a uniform magnetic field, $H$, under certain
conditions, also can  induce a gap in the excitation spectrum of
1/2HAFC. While it is well established that the ground state of a
1/2HAFC in a magnetic field  remains gapless provided that $H$
does not exceed the saturation field,  Dender et al.\cite{dender},
performing neutron scattering experiments on the 1D
antiferromagnetic chain-like compound Cu benzoate, have recently
discovered that the magnetic field does create a gap in the spin
excitations. They suggested that a staggered field $h$, arisen
from the staggered $g^s$-tensor associated with the low local
symmetry of the Cu$^{2+}$ paramagnetic centers, is at the origin
of this unexpected behavior. During the last years, these findings
have generated a great deal of activity on a 1/2HAFC in a
staggered field \cite{oshikawa97,essler97,essler98,affleck99}.

The electron spin resonance (ESR) is a powerful tool for probing
the low-energy magnetic excitations in solids with a unique
sensitivity and resolution.
%So it is rather tempting to apply the
%ESR for studying spin excitations in 1/2HAFC.
However, a clear lack of a rigorous ESR theory, until very
recently, has seriously retarded carrying out  such experiments on
1/2HAFC . Notice that in conventional ESR theories the most
important ingredient of the ESR physics, the dynamical four-spin
correlation function, is evaluated  within various decoupling
schemes \cite{Kubo,Mori}. Naturally, these kinds of approach fail
in low-dimensional quantum spin systems where the many-body
correlation effects are strong.

It was not until 1999 that Oshikawa and Affleck (OA) have
developed an ESR theory for a 1/2HAFC based on a field theory
approach \cite{oshikawa99,oshikawa02}.
%They computed the dynamical
%correlation functions  within a well controlled approximation
%early proposed by Schulz \cite{schulz86}.
As the perturbations of
1/2HAFC, ensuring  a nonzero linewidth, OA considered  a staggered
field $h$ caused by the staggered $g^s$-tensor and the
Dzyaloshinskii-Moriya (DM)interaction, $\Sigma \vec {D}\cdot \vec{
S}_{l}\times \vec{ S}_{l+1}$.They have shown that in a weak
magnetic field, $h$ can be approximated  by $\vec h \approx
g^s\vec{ H}+\vec {D}\times g^u\vec {H}/2J$, $J$ being an
intrachain exchange \cite{oshikawa97}.  Let us summarize briefly
their  main results as far as the ESR linewidth, $\Delta H$,  is
concerned (for a more detailed discussion, see below). In the
absence of  $h$, the predictions of OA's theory are remarkably
simple. The linewidth of  the ESR signal in 1/2HAFC  is linearly
proportional to $T$ for temperatures  $T\ll J$, exhibiting a weak
angular dependence on the direction of $H$. Now if 1/2HAFC is
subjected to a staggered field $h$,   the behavior of $\Delta H$
drastically changes: instead of having
 $\Delta H\sim T$  valid for $h=0$, the width acquires a peculiar low-$T$
dependence on $h$ and $T$, $\Delta H \sim (h/T)^2$, diverging at
low temperature. Experimentally, the magnitude of a staggered
field can be controlled by a rotation of the applied magnetic
field $H$. In the case where $h$ is mainly due to the DM term,
$h=0$ if $H\| D$ and $h=h_{max} $ for $ H\bot  D$, as it is easily
seen from the above expression for $\vec h$. Since $h \sim H$,
this should lead to a strong dependence of $\Delta H$ on the $H$
direction. Finally, at very low temperature, $T\ll H$, the ESR
spectrum is dominated by soliton-like excitations
\cite{oshikawa99}.

Some of these theoretical results have already received  support
through  ESR experiments on Cu benzoate. In particular, the very
low-$T$ regime has been studied in detail in several works
\cite{asano,sakon}. There remain, however, interesting open
problems related to the low-$T$  and intermediate-$T$ spin
dynamics of 1/2HAFC. To date, to the best of our knowledge, there
has been no convincing evidence confirming the $T$-linear
dependence of $\Delta H$ in a 1/2HAFC at intermediate
temperatures. As far as the low-$T$ $(h/T)^2$- contribution to
$\Delta H$ is concerned, although the  ESR data on Cu benzoate are
found to be in  good qualitative agreement  with the theoretical
predictions, a quantitative comparison between the theory and the
experiment
 is still lacking.
Even more importantly,  the only material which has been studied,
up to now, for the purpose of testing the OA ESR theory is Cu
benzoate. Clearly, new model 1/2HAFC systems are necessary in
order to carry out a comprehensive test of the theoretical
predictions. In this Letter   all of these points are addressed by
performing an extensive ESR study in a wide frequency and
temperature range on a new $S=1/2$ chain-like antiferromagnet with
DM interaction, BaCu$_2$Ge$_2$O$_7$. Our results provide new
evidence in  favor of the OA theory.

The crystal structure of BaCu$_2$Ge$_2$O$_7$,which  belongs to the
orthorhombic space group \textit{Pnma} ($Z=4$),  is made of zigzag
chains of corner-sharing CuO$_4$ plaquettes running along the
crystallographic \textit{c} axis (Fig.\ 1a) ( for  a detailed
description of the crystal structure, see Ref.\ \cite{tsukada99}).
The Cu-O-Cu bond angle is found to be 135$^{\circ}$. This is much
smaller than the 180$^{\circ}$-angle that occurs in superexchange
bridges of many 2D cuprates. This implies that the staggered
$g^s$-tensor and the DM interaction are present in the spin
Hamiltonian of an individual CuO$_3$ chain. The magnetic
properties of BaCu$_2$Ge$_2$O$_7$ are known from the
neutron-diffraction, magnetic susceptibility, $\chi (T)$,  and
magnetization measurements \cite{tsukada99,tsukada00}. From the
low-$T$ part of $\chi (T)$  measured below  350 K, the intrachain
exchange integral $J\approx 540$ K has been extracted by
comparison with the Bonner-Fisher curve for a 1/2HAFC. Interchain
interactions in BaCu$_2$Ge$_2$O$_7$ lead to an antiferromagnetic
long-range order at low $T$, $T_N =8.8$ K, with spins aligned
parallel to an easy $c$-axis and a weak ferromagnetic moment along
$b$-axis as established  in the magnetization measurements
\cite{tsukada00}. Thus this compound, with a ratio $T_N/J=0.017$,\
can be viewed  as a nearly ideal model system for a 1/2HAFC with
the DM interaction.

The single crystals of BaCu$_2$Ge$_2$O$_7$ used in the
 ESR experiments were grown by a
floating-zone method. Particular attention was paid to the crystal
orientation, since the $a$- and $c$-axis lengths are almost the
same. The ESR spectra in X-band ($\nu =9.6$ GHz) and Q-band ($\nu
=34$ GHz) were collected using a Bruker EMX spectrometer. At high
frequencies, a homemade millimeter-range video spectrometer  has
been used.

In order to minimize the effect of the $(h/T)^2$-contribution to
$\Delta H(T)$ which can be important at low $T$, the ESR at the
lowest frequency (lowest resonance magnetic field)  was performed
first. A broad single ESR line was  characterized by the following
ESR parameters at room temperature: $\Delta H_a=0.2650\pm 0.0002$
T, $\Delta H_b=0.1990\pm 0.0002$ T, $\Delta H_c=0.1740 \pm 0.0002$
T and $g_a=2.22\pm 0.001$, $g_b=2.08\pm 0.001$, $g_c=2.09 \pm
0.001$, where the subscripts $a,b,c$ indicate the magnetic field
direction.

The temperature  dependence of the  linewidth in
BaCu$_2$Ge$_2$O$_7$ for magnetic field orientation along the three
principal axes at $\nu =9.6$ GHz is shown in Fig.\ 1b. For
comparison $\Delta H(T)$ data for KCuF$_3$, to date the most
studied 1/2HAFC-like material, from Ref.\ \cite{yamada}, are also
presented. Please note that normalized coordinates are used both
for $\Delta H$  and the temperature (the latter is normalized by
$T_{max}\approx 0.64 J$, the temperature corresponding to the
maximum of the static susceptibility). The values of  $T_{max}
=346 $\ K and $T_{max} =243$\ K were taken for BaCu$_2$Ge$_2$O$_7$
and for KCuF$_3$, respectively. $\Delta H$ is normalized by its
value at $T=T_{max}/2$. As is clearly seen from this figure, when
$T$ decreases and approaches  $\sim T_{max}/2$, the slope of
$\Delta H(T)$ changes, displaying a crossover
 from the "classical" Kubo-Tomita regime to  the predicted
quantum one. At this temperature the correlation length along the
chains $\xi/a \approx 2$ (where $a$ is the distance between
nearest Cu sites in the chain direction). This quite naturally
establishes the limits of the validity of the field theory
approach. Below $ T_{max}/2$, the linewidth, indeed, follows the
linear dependence $\Delta H\sim T$ in a large $T$-interval, as
predicted by OA's theory (see Ref. \cite{comment} for alternative
models) .

At low $T$ another contribution to $\Delta H$,  which we  call
$\Delta H_{3D}$, becomes important. Since $\Delta H_{3D}$
contributes to $\Delta H$ at temperatures close to $T_N$ we
tentatively interpret it as due to the 3D N\'{e}el phase
fluctuations. As $T$ decreases, $\Delta H_{3D}$ first increases
and then shows a tendency to decrease. We have tried to fit
$\Delta H(T)$ for $H\|c$ using the equation $\Delta H(T)\sim
A\cdot T+\Delta H_{3D}(T)$, where a diverging part of $\Delta
H_{3D}$ is represented by $\Delta H_{3D}(T)=B\cdot(T-T_N)^n $, in
order to have an  estimate for $n$. The fit, shown in  Fig.\ 1b by
the solid line, gives $n=-1.1$. Thus, $\Delta H_{3D}$,  can not be
identified with the $(h/T)^2$-contribution and, in fact, as we
show below, is found not to depend on the magnetic field.

Let us now turn to our results of high-frequency ESR experiments
on BaCu$_2$Ge$_2$O$_7$. In what follows we shall compare
systematically the ESR data for two field orientations, $H\|a$ and
$H\|c$. Fig.\ 2a shows the linewidth, as a function of temperature
below $70$ K for $H\|a$, measured at frequencies 9.6, 34 and 134
GHz. Note that in this $T$ interval the above discussed $\Delta
H_{3D}(T)$ contribution is of main importance. Despite the fact
that the operating frequency is changed almost by a factor of 15,
$\Delta H(T)$ remains unchanged within experimental accuracy.
Therefore it is  reasonable to assume that the sum of the linear
and $\Delta H_{3D}$ contribution is frequency independent. In
Fig.\ 2b we have plotted  $\Delta H(T)$ for $H\|c$ as the
difference between $\Delta H(T,\nu)$ measured at the frequency
$\nu$ ($\nu=$ 34, 70.6 and 134 GHz) and $\Delta H(T,9.6\
\textrm{GHz})$ measured at $\nu=9.6\ \textrm{GHz}$. (Notice that
the maximum value of $\Delta H(T,9.6\  \textrm{GHz})$ at low-$T$,
i.e. the value that is subtracted from the shown curves, is less
than 0.13 T.) The ESR data for the magnetic field direction $H\|c$
is found to be  in significant contrast with those of $H\|a$. As
evident from Fig.\ 2b, $\Delta H(T,\nu)$ strongly depends on both
the temperature and the frequency (or the resonance field).
According to OA's theory this is a staggered field, $h$, induced
by an applied uniform magnetic field $H$ which is responsible for
these phenomena. Therefore, to compare the obtained results with
the theory one needs more information about $h$. To this end we
have undertaken antiferromagnetic resonance (AFMR) studies on
BaCu$_2$Ge$_2$O$_7$ below $T_N$.  The idea to use the AFMR for the
staggered field determination resides in the fact that, for $H$
applied along the principal crystal axes, among the two resonance
modes there exists one, called "antiferromagnetic mode", which
depends strongly on $h$, $\nu \sim \sqrt{h} $ and another one,
usually called "ferromagnetic" mode, which depends mostly on $H$,
$\nu \sim H$ \cite{turov}.

Experimentally, two modes of  antiferromagnetic resonance have
been  found in BaCu$_2$Ge$_2$O$_7$ at $2$\ K. The corresponding
frequencies at $H=0$ , $\nu_{1}(0)=22$ GHz and $\nu_{2}(0)=56$
GHz. What we need in order to proceed in our discussion of the
staggered field in BaCu$_2$Ge$_2$O$_7$  is the magnetic field
dependence of $\nu_{2}$ for $H\|c$ and that of $\nu_{1}$ for
$H\|a$. In Fig.\ 3 $\nu_{2}(H)$ for $H\|c$ measured at $T=2$\ K is
shown. The solid line follows the  equation
$(\nu_2/\gamma)^2=(\nu_{20}/\gamma)^2+ 4SJh$ which results from
the spin-wave theory for a 1/2HAFC, where  the staggered field is
given by  $h=C\cdot H$ (see, for example, \cite{affleck99}). The
best fit gives a product $4SJ\cdot C=18.7\pm 0.15$ T. Since $S$ is
strongly reduced by the zero-point fluctuations, to get $C$, an
estimate of the mean field value of $S$, $\overline{S}$, is now
required. This can be done by means of a recent theory by Irkhin
and Katanin \cite{irkhin}.
%which treats the magnetic properties of
%weakly interacting 1/2HAFC within the bosonization method.
Inserting the values of $J=540$ K and $T_N=8.8$ K in their
eq.(21), one obtains $\overline{S}=0.088$ which leads to $C=0.14$
for $H\|c$. As far as the field dependence of  $\nu_{1}$ for
$H\|a$ is concerned we have found that $\nu_{1}(H)$ varies only
slightly with $H$. Nevertheless, $4SJ\cdot C$ can be estimated in
this case as
 $ 1.2\pm 0.25$T leading to $C=0.0074$. [Note, however, that an
imperfect magnetic field orientation  could also be at the origin
of this small variation of $\nu_{1}(H)$.] Thus, our  important
conclusion, which is due to the high frequency AFMR measurements,
is that the magnitude of  $h$ induced by the applied magnetic
field $H$ strongly depends on the direction of $H$, taking a
maximum value for $H\|c$. This is consistent with the ESR
linewidth data presented in Fig.\ 2\ : $\Delta H$ is  frequency
independent for $H\|a$ and it exhibits a rather strong frequency
(magnetic field) dependence for $H\|c$.

We are now at a point to make a quantitative comparison between
our results and those of OA's theory. Their prediction for the
temperature and magnetic field  linewidth dependence  at low $T$
is $\Delta H\approx 0.69($ln$J/T)Jh^2/T^2$. This equation, with
the logarithmic factor replaced by unity, was already used to
describe the ESR data on copper benzoate
\cite{oshikawa99,oshikawa02}. Therefore we try first to fit our
data in Fig.\ 2b making the same assumption. The best fit giving
$C=0.16$ is shown in Fig. 2b by the solid lines. Comparing this
with our estimate from the AFMR experiment, $C=0.14$, one can
conclude that the OA equation, taken with ln$(J/T)\approx 1$,
slightly underestimates $\Delta H$, by approximately $ 15\%$, but
that it captures the essential of the spin dynamics of 1/2HAFC in
a staggered field at a quantitative level. Let us now test the
above
 equation in its complete form (i.e. including the log
correction). As it can be seen from Fig. 2b, where the theoretical
curves are presented by the dashed lines, again the equation fits
well to the experimental data giving $C=0.09$, which is about $35
\%$ less than the experimental value. Since there are no
adjustable parameters in the theory, other than $C$, one should
consider the theory to be in excellent agreement with the
experiment.

In conclusion, we have undertaken  detailed ESR measurements on
the new $S$=1/2 quasi-one dimensional compound with
Dzyaloshinskii-Moriya interaction, BaCu$_2$Ge$_2$O$_7$.
%, in order
%to test the recent theory of Oshikawa and Affleck describing the
%ESR in such systems.
We have essentially focused on the temperature and frequency
behavior of the linewidth in two distinct $T$-intervals. At
intermediate $T$, in accordance with the OA theory, the linewidth
is found to be linearly proportional to the temperature and
frequency independent, regardless of the field orientation. The
crossover from the Kubo-Tomita behavior to the $T$-linear regime
takes place at around $T_{max}/2 $. This establishes the high-$T$
limit of the validity of the field theory approach. At low
temperatures ($T_N< T\ll T_{max}/2$) the linewidth behavior
strongly depends on the magnetic field orientation. For $H\| c$,
 corresponding to the maximum of the staggered field $h$
induced in the system, the linewidth scales approximately  as $ (
h/T)^2$ in agreement with the theory. To infer the key theory
parameter, the proportionality constant $C$, from an independent
experiment, we have carried out AFMR measurements at $T < T_N$. We
were able then, for the first time, to perform a detailed
quantitative comparison between our results and the theoretical
predictions.

\newpage
%FFFFFFFFFFFFFFFFFFFFFFFFFFFFFFFFFFFFF    FIGURE 1
\begin{figure}[t]
\includegraphics[bb= 22 620 505 770,clip,width=0.7\textwidth,angle=0]{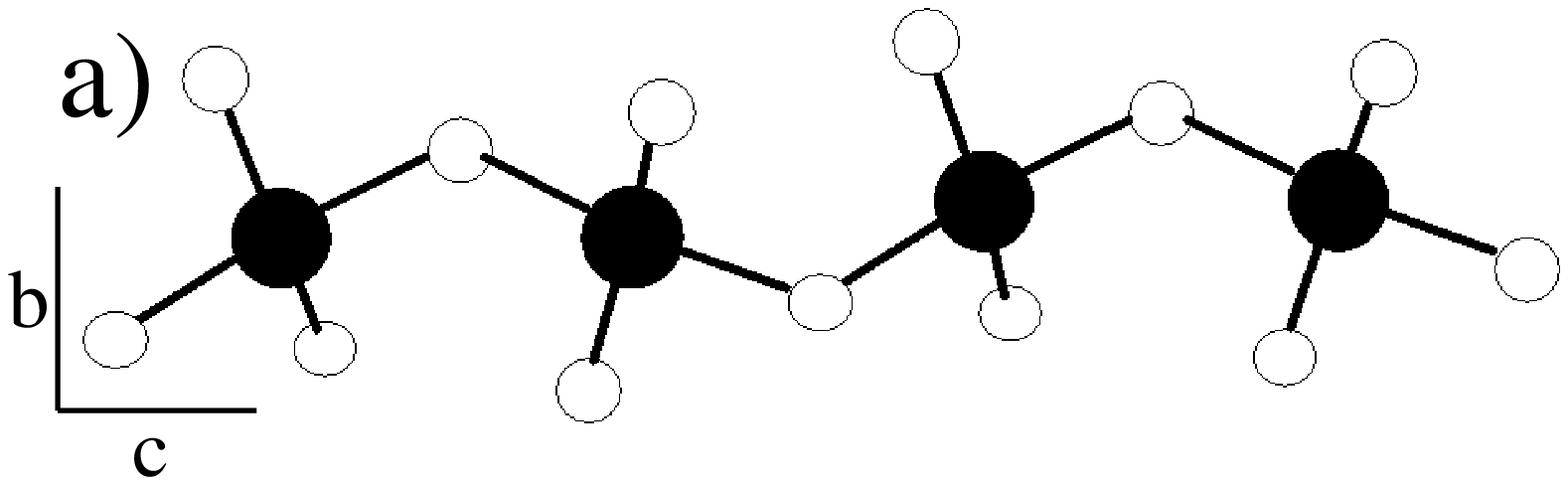}
\includegraphics[width=0.7\textwidth,angle=0]{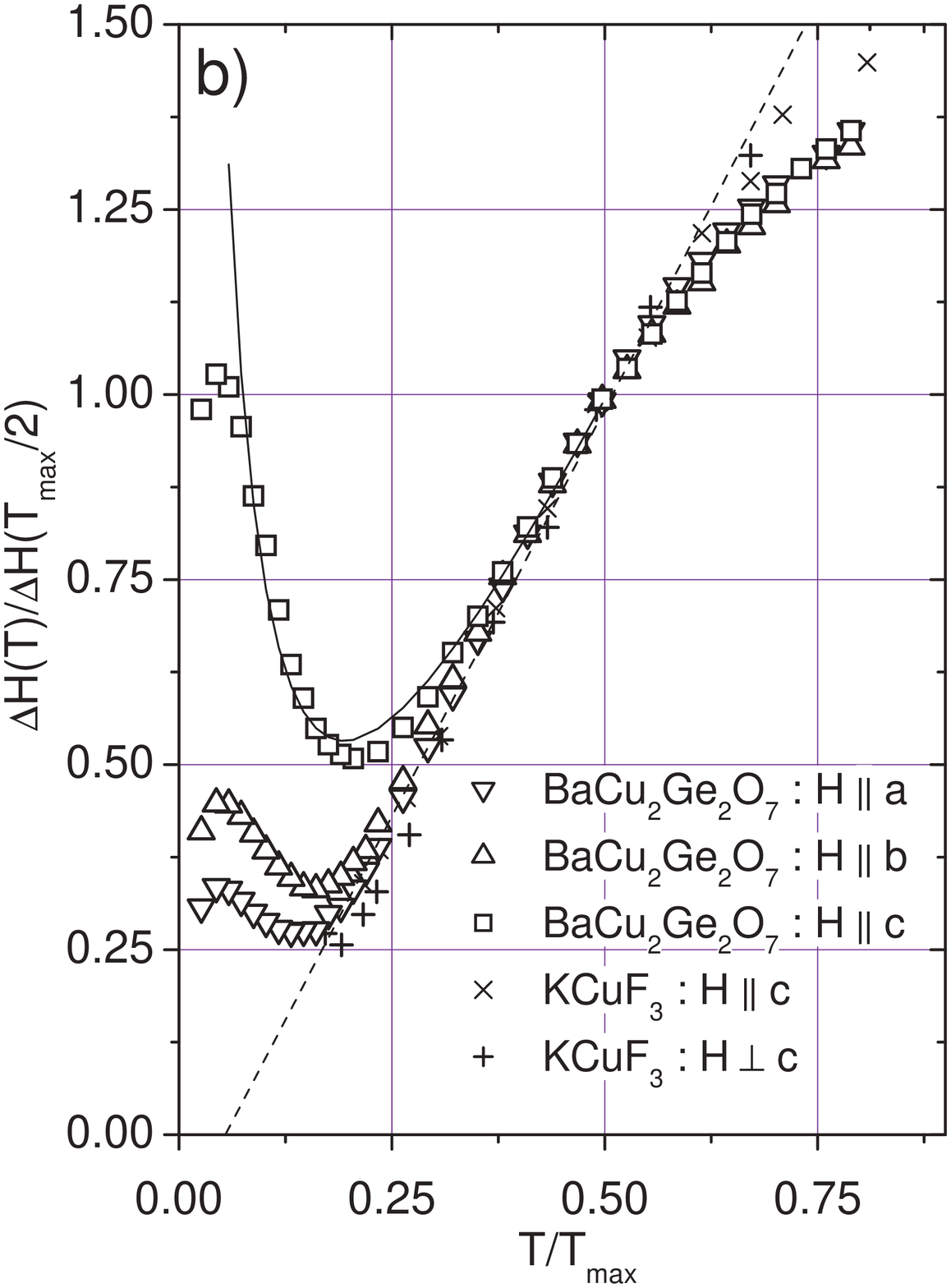}
\caption{ (a) The (b,c)-plan projection of a fragment of the zigzag
chain of corner-sharing CuO$_4$ plaquettes. The black circles are
Cu$^{2+}$ ions, the open circles are O$^{2-}$; (b) The temperature
dependence of the linewidth in BaCu$_2$Ge$_2$O$_7$ along the three
principal crystallographic directions, at $\nu=$9.6  GHz, presented
in normalized coordinates (see text). Data for KCuF$_3$ from Ref.\
\cite{yamada}  are also shown.  The solid line results from the fit,
the dashed line is  a guide to the eye. }
\end{figure}

\newpage
%FFFFFFFFFFFFFFFFFFFFFFFFFFFFF FIGURE 2

\begin{figure}[b]
\includegraphics[width=.8\textwidth]{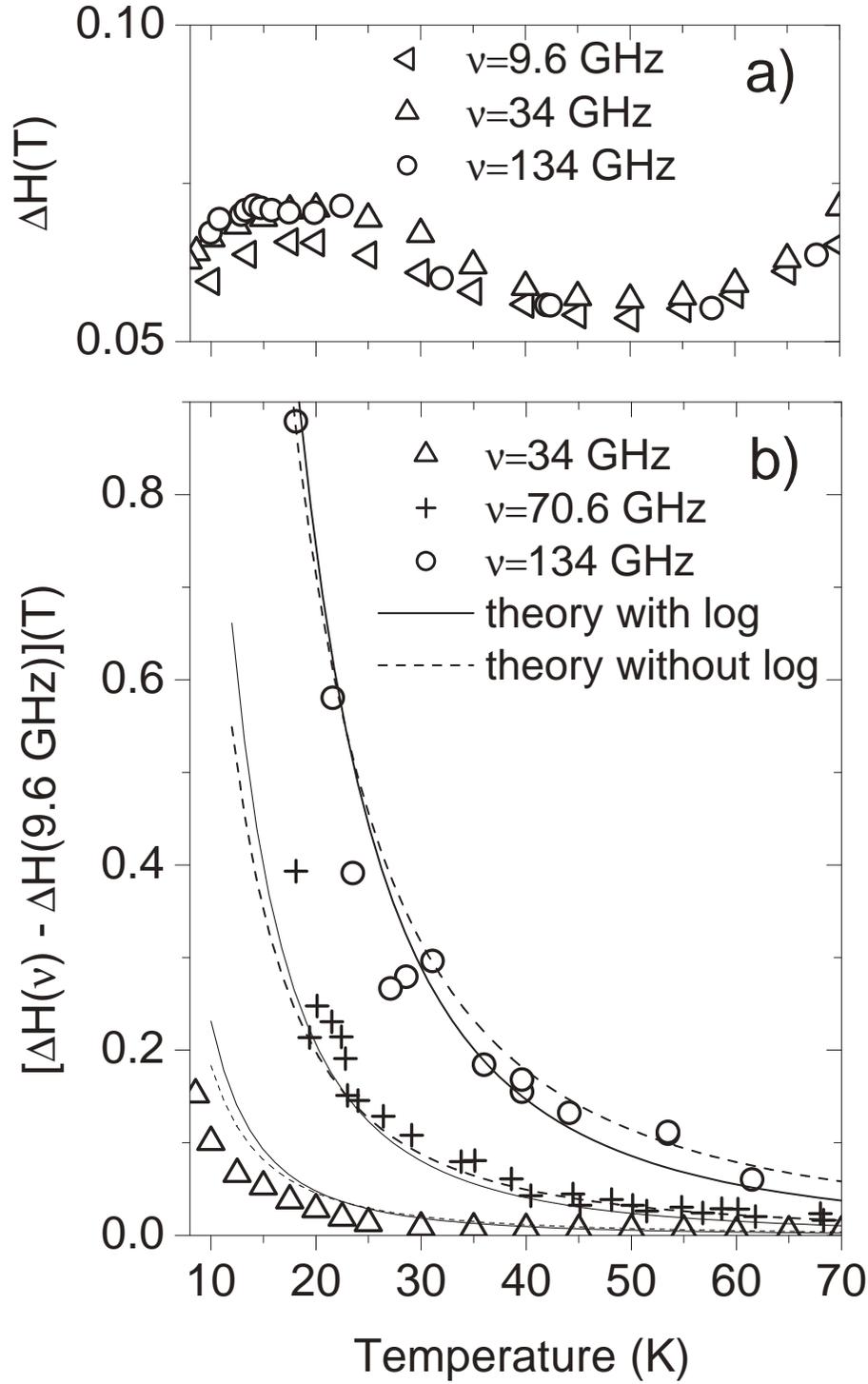}
\caption{ a) $\Delta H(T)$ for $H\|a$ as a function of frequency; b)
The temperature dependence of the difference between $\Delta
H(T,\nu)$ measured at a frequency $\nu$ and $\Delta H(T,9.6\
\textrm{GHz})$
 measured at $\nu=9.6$ GHz, for $H\|c$ in BaCu$_2$Ge$_2$O$_7$. The solid and the
dashed lines are according to Eq.\ (9) of Ref.\ \cite{oshikawa99}. }
\end{figure}

\newpage

%FFFFFFFFFFFFFFFFFFFFFFFFFFFFF FIGURE 3

\begin{figure}[tbt]
\includegraphics[width=.8\textwidth]{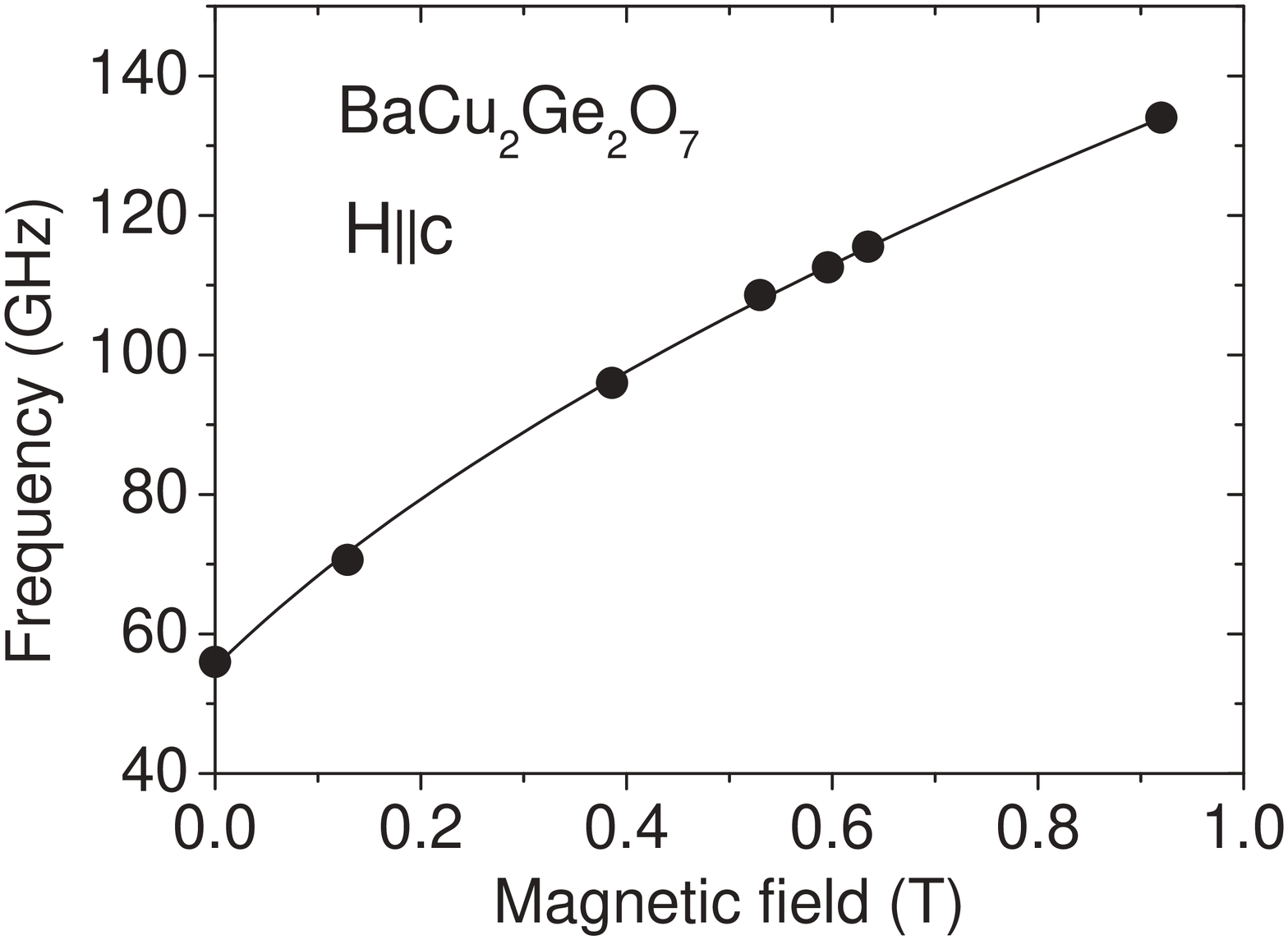}
\caption{ The frequency-vs-field diagram of  the AFMR mode
$\nu_2(H)$ in BaCu$_2$Ge$_2$O$_7$ at $T=2$K, for $H\|c$.}
\end{figure}

\newpage


\begin{references}

\bibitem[\S] {SB} email : sylvain.bertaina@l2mp.fr

\bibitem[\dag] {TM}Present address: Condensed Matter Sciences Division, Oak Ridge
 National Laboratory,
 Oak Ridge, Tennessee 37831-6393, USA.

\bibitem[\ddag]{KU}Present address: RIKEN (The Institute of Physical and Chemical
  Research), Wako 351-0198, Japan.
\bibitem{bethe}
H. A. Bethe, Z. Phys. {\bf 71}, 205 (1931).

\bibitem{hase}  M. Hase et al., Phys. Rev. Lett.
{\bf 70}, 3651 (1993).

\bibitem{dender} D. C. Dender et al., Phys. Rev. Lett. {\bf 79}, 1750 (1997).
\bibitem{oshikawa97}%field induced
M. Oshikawa and I. Affleck, Phys. Rev. Lett. {\bf 79}, 2883
(1997).
\bibitem{essler97}F. H. L. Essler et al.,\
Phys. Rev. B {\bf 56},\ 11001 (1997).

\bibitem{essler98} F. H. L. Essler and A. M. Tsvelik,\ Phys. Rev. B {\bf 57}, 10592 (1998).

\bibitem{affleck99}
I. Affleck and M. Oshikawa, Phys. Rev. B {\bf 60}, 1038 (1999).

\bibitem{Kubo}  R. Kubo and K.Tomita, J. Phys. Soc. Jpn. {\bf 9}, 888 (1954).

\bibitem{Mori}  H. Mori, Progr. Theor. Phys., {\bf 34}, 423
(1965).

\bibitem{oshikawa99}%ESR
M. \ Oshikawa and I.\ Affleck, Phys.\ Rev.\ Lett. {\bf 82}, 5136
(1999).
\bibitem{oshikawa02}%ESR
M.\ Oshikawa and I.\ Affleck, Phys.\ Rev.\ B {\bf 65}, 134410
(2002).

\bibitem{asano}
T.\ Asano et al., Phys.\ Rev.\ Lett. {\bf 84}, 5880 (2000).

\bibitem{sakon}
T.\ Sakon et al., J. Phys. Soc. Jpn. {\bf 70}, 2259 (2001).

\bibitem{tsukada99}
I.\ Tsukada et al., Phys.\ Rev.\ B {\bf 60}, 6601 (1999).

\bibitem{tsukada00}
I.\ Tsukada et al., Phys.\ Rev.\ B {\bf 62}, R6061 (2000).


\bibitem{yamada}
I.\ Yamada, et al.,\ J.\ Phys.:\ Condens. \ Matter {\bf 1}, 3397
(1989).
\bibitem{comment}
Various models were proposed in order to explain the linear-T
dependence of $\Delta H$, $\Delta H(T)=\alpha+\beta T$. These are
based either on the spin diffusion concept \cite{richards},  the
spin-phonon coupling mechanism \cite{seehra} or the considering of
the effect of static spin correlations on the linewidth
\cite{soos}. However, all of these models are valid at relatively
high temperatures, $T>J$, as compared with the $T$ interval which
is of interest in the present study. Besides, note that the latter
model suffers from an improper evaluation of the second moment in
the case of the 1/2HAFC with the DM interaction
\cite{oshikawa02,choukroun}.

\bibitem{richards}
P.\ M.\ Richards in {\em Local Properties at Phase Transitions,
edited by K. A. M\"{u}ller}. Amsterdam, North-Holland Publishing
Company (1976).


\bibitem{seehra}
M. S. Seehra and T. C. Castner, Phys. Kondens. Materie {\bf 7},
185 (1968).

\bibitem{soos}
Z. G. Soos et al., Chem. Phys. Lett. {\bf 46}, 600 (1977).

\bibitem{choukroun}
J. Choukroun et al., Phys. Rev. Lett.  {\bf 87}, 127207 (2001).

\bibitem{turov} E. A. Turov in {\em  Physical
properties of magnetically ordered crystals}. New York, Academic
Press (1965).



%\bibitem{bertaina2003}

%S.\ Bertaina et al.  (unpublished).


\bibitem{irkhin}
V.\ Yu.\ Irkhin and A.\ A.\ Katanin, Phys. Rev. B {\bf 61}, 6757
(2000).




\end{references}
\end{document}